\documentclass{JHEP3}
\usepackage{epsfig}
\usepackage{graphicx}% Include figure files
\usepackage{dcolumn}% Align table columns on decimal point
\usepackage{bm}% bold math
\newcommand{\be}{\begin{equation}}
\newcommand{\ee}{\end{equation}}
\def\bea{\begin{eqnarray}}
\def\eea{\end{eqnarray}}

 \newcommand{\Ka}{K\"ahler}
\def\lesssim{\mathrel{\hbox{\rlap{\hbox{\lower4pt\hbox{$\sim$}}}\hbox{$<$}}}}
\def\gtrsim{\mathrel{\hbox{\rlap{\hbox{\lower4pt\hbox{$\sim$}}}\hbox{$>$}}}}

\title{On Brane Inflation With Volume Stabilization}

\author{Jonathan P. Hsu, Renata Kallosh  and Sergey Prokushkin
\\

Department of Physics, Stanford University,
\\ Stanford, CA 94305-4060, USA.\\
 e-mail: pihsu@stanford.edu, kallosh@stanford.edu, prok@stanford.edu}

\received{November  2003}       %%
 \preprint{SU-ITP-03/26\\  ~hep-th/0311077\\  November  2003}

\abstract{\small The distance between BPS branes in string theory corresponds to a flat direction in the effective potential. Small deviations from supersymmetry may lead to a small uplifting of this flat direction and to brane inflation. However, this scenario can work only if the BPS properties of the branes and the corresponding flatness of the inflaton potential are preserved in  the stable volume compactification. We present an ``inflaton trench'' mechanism that keeps the inflaton potential flat due to shift symmetry, which is  related to 
 near BPS symmetry in our model. 
 }
 
\begin{document}
%%%%%%%%%%%%%%%%
%\tableofcontents{}

\parskip 5pt
\section{Introduction}

It is difficult to find a mechanism for inflation in string
theory. Brane inflation models
\cite{dbraneinflation,Herdeiro:2001zb,Dasgupta:2002ew,angles}
provide a significant step towards this goal. However, the
validity of  brane inflation models in string theory depends on
the ability to
stabilize the compactification volume. This means that an
effective 4d theory has to be found in which there is a potential
for the scalar field representing the volume of the internal
space. This potential has to have a minimum in which the volume
field is trapped. This potential should simultaneously describe
the inflaton field corresponding to the distance between branes.
The
effective theory has to fix the volume modulus while keeping the
potential  for the distance modulus flat.

Brane inflation models preserving
supersymmetry should not have any potential for the modulus
representing the distance between BPS branes. When supersymmetry
is slightly broken, so that the configuration of branes is a near
BPS state, a small attractive potential develops for the distance
modulus, which in turn may lead to a slow-roll stage of inflation
in the effective 4d theory. Brane inflation models of this type,
where small deviations from the BPS configuration trigger
inflation were studied in
\cite{Herdeiro:2001zb,Dasgupta:2002ew,angles}.

In the effective field theory, this situation may be
characterized by a property of the inflaton field: its mass should
be small and definitely much smaller than the Hubble parameter,
$m^2_{\rm infl}\ll H^2$. In supergravity there are known
mechanisms to provide this type of relation. Brane inflation models
developed without an explicit moduli stabilization mechanism
assumed that the stabilization mechanism would not lead to a large
inflaton mass. However, the recent study of D3/D$\bar{3}$
inflation in a warped background with fluxes and stable volume
modulus \cite{Kachru:2003sx} has shown that the 
%old 
danger of
$m^2_{\rm infl}\sim H^2$ may persist, unless a fine-tuning can be performed.

The purpose of this paper is to suggest a mechanism which protects the inflaton field from
acquiring a large mass in the context
of stable volume brane inflation models. The mechanism can be intuitively understood as
exhibiting the fact that BPS branes should be in neutral
equilibrium. Compactification with a stable volume should preserve
this fact under certain conditions.

We originally found this mechanism by trying  various potentials
in Mathematica. We were looking for the possibility of having some
kind of a flat direction: the inflaton should form a trench which gives an
additional dimension to the narrow well trapping the volume modulus
in the simplest KKLT model \cite{Kachru:2003aw} or in more general
models \cite{egq}. We have dubbed the phenomena as the ``inflaton trench''
mechanism. Only later did we realize that the corresponding
potentials allow the realization of an {\it exact unbroken
supersymmetry} in AdS space. One may try to keep this property for
near BPS branes in near dS space, where it will be only
approximately correct, but will preserve
the $m^2_{\rm infl}\ll H^2$ relation.

As a specific example, we will work with the D3/D7 brane
inflation model which  is related to the hybrid model
\cite{Linde:1991km} of D-term inflation \cite{Binetruy:1996xj}. We
will find that it is compatible with the KKLT mechanism of the
volume stabilization. D-term inflation models in supergravity do
not require a fine-tuning with regard to the inflaton mass since
the source of inflationary energy is coming from the D-term while the
F-term vanishes during inflation. Therefore the dangerous \Ka\
corrections, which could lead to $m^2_{\rm infl}\sim H^2$, vanish. However, the superpotential volume stabilization mechanism suggested in the KKLT model
\cite{Kachru:2003aw} requires some non-vanishing F-terms. In this paper we will suggest
a consistent combination of these two  ideas which may provide the basis for stringy brane
inflation models with a stable volume.

\section{D3 branes, D7 branes and the volume modulus}

It has been pointed out in \cite{Kachru:2003sx} that, in brane cosmologies, it is
important to identify  properly the volume modulus and the
modulus responsible for the distance between
branes. For
compactifications with only D3 and D$\bar{3}$ branes present,
there are arguments that the physical volume of the
internal space is given by the combination of the real part of the
chiral superfield $\rho+\bar \rho$ and the \Ka\ potential $k$ of
the Calabi-Yau metric. 
\be 2 r= \rho+\bar \rho -k(\phi,
\bar \phi)\, .\label{volumeD3} \ee 
One of the arguments presented in
\cite{DeWolfe:2002nn} is that the kinetic term for the coordinates
of the D3 brane $\phi$ has the form 
\be 
T_{D3} \int d^4 x
{\partial_\mu \phi^i \partial^\mu \bar \phi^{\bar k} k_{i\bar
k}\over \rho+\bar \rho} \, . 
\label{kineticD3}\ee
 This in turn means
that the \Ka\ potential of the effective 4d theory must be \be
K(\rho, \phi, \bar \rho, \bar \phi)_{D3}= -3 \log (\rho+\bar \rho
-k(\phi, \bar \phi))\ .\label{kahlerD3} \ee 
At small $|\phi|^2$, $k(\phi, \bar \phi)=
\phi\bar \phi+ \dots $, and one finds that, generically without fine-tuning, 
in models of inflation 
with D3 branes the inflaton mass
problem, $m^2\sim H^2$, persists.

In the case of a compactification with
only D7 branes present it was pointed out in
\cite{Burgess:2003ic} that, by dimensional reduction,
the real part of the volume modulus appears in the gauge coupling
constant 
\be {1\over g^2} (F_{\mu\nu})^2\sim (\rho+\bar \rho)
(F_{\mu\nu})^2 \, . \ee 
Since the
gauge coupling has to be given by the real part of the chiral
superfield the dependence on $S$, the coordinate of
the D7,  should not enter in the definition of the volume
modulus in the form $\rho+\bar \rho -k(S, \bar S)$, we should have
 \be 2 r= \rho+\bar \rho \, . \label{volumeD3a} \ee 
 An additional argument for identifying the physical volume
in this fashion is provided by the calculation of the kinetic
term for the D7 coordinates, $S$, which gives 
\be T_{D7} \int d^8 x
\partial_\mu S^i \partial^\mu \bar S^{\bar k} k_{i\bar k}
\label{kineticD7} \, .\ee 
This suggests that the \Ka\ potential
must be given by 
\be K(\rho, S, \bar \rho, \bar S)_{D7}= -3 \log
(\rho+\bar \rho) +k(S, \bar S) \label{kahlerD7} \, .\ee 
For small
$|S|$, or in some specific models, $k(S, \bar S)\sim
S\bar S+\dots$. In the setting of \cite{Dasgupta:2002ew} the type IIB string theory is compactified on $K_3\times {T^2\over \mathbb{Z}_2}$. The torus is oriented along $x^4, x^5$. Therefore in these directions, for $S\sim x^4+i x^5$ we have a canonical metric with $K=S\bar S$.
 The
distance between branes should be given by
the difference between the coordinates of D3 and D7. Therefore
there is a puzzle: what is the correct \Ka\ potential when both D3
and D7 branes are present, as in the brane construction for the
D-term inflation model \cite{Dasgupta:2002ew}. How do we resolve
the contradiction between these two expressions for
$K$?

We propose to resolve this contradiction as follows. If one
object is much heavier than the other object, the lighter object should
be thought of as a probe which does not significantly modify the
background.

The case of the D3/D7 system may be considered as a D7 brane
probing the geometry of a heavy stack of D3
branes. If there are many coincident D3 branes, one may treat a D7
brane with fluxes as a probe of the D3 background. This was in
fact done in \cite{Dasgupta:2002ew} in sec. 2.2 where the D7 brane
world-volume action in a D3 background was studied.

In a situation when the D7 is probing the geometry defined by the
stack of D3 branes, the relative distance between the D7 and the
stack of D3 branes is defined by the coordinates of the D7. Since
the heavy D3s do not move, we may use an effective field theory
with the \Ka\ potential of the type defined by the D7 worldvolume
in eq. (\ref{kahlerD7}).  We will consider the position of the D3
brane, $\phi$, not as a dynamical variable. However, $S$, being
the position of the D7 will be treated as a dynamical variable.
Thus there are two different physical systems: when the D7 is
considered as a probe, the \Ka\ potential of the effective field
theory is given by (\ref{kahlerD7}). When the D3 brane is
considered as a probe, the \Ka\ potential of the effective field
theory is given by (\ref{kahlerD3}). We believe that this
distinction allows us to resolve the puzzle\footnote{The existence
of the puzzle was revealed in discussions with C. Burgess, J.
Maldacena, S. Kachru, S. Trivedi and F. Quevedo.} of having two
different choices of the \Ka\ potential.

The analysis of the Kahler potentials and the significant difference 
 between the coordinates describing the position of the D7 brane and D3 
 brane suggested here can  also  be understood in the framework of  the 
 special geometry of the modulli space \footnote{We are grateful to S. 
 Ferrara and C. Angelantonj  for the explanation of this issue.}  as derived for type IIB  string 
 theory compactifications in \cite{Andrianopoli:2003jf} .

\section{BPS branes in AdS space with volume stabilization}

Consider some combination of branes with fluxes which form a BPS state with some number of
unbroken supersymmetries. For example, consider a D7 brane placed
in the $0,1,2,3,6,7,8,9$ directions with a self-dual flux on its
worldvolume in the compact directions, $F_{mn}=
F_{mn}^*, m,n=6,7,8,9$ . Add to this
system many D3 branes extended in the $0,1,2,3$ directions at
some distance from the D7 in $4,5$ directions. This system is
supersymmetric. The distance between the branes, $(x^4)^2+
(x^5)^2$, can take any value without changing the fact that this
is a BPS state. There is no force between branes, or, to be more
precise, there is a balance of gravitational, dilaton and RR
form-field forces.

The condition for unbroken supersymmetry for a D7
brane in the background geometry of a D3 brane with a harmonic
function $H(\vec y^2)$ was defined in \cite{Dasgupta:2002ew} by
solving the equation \cite{Bergshoeff:1997kr}
\be (1-\Gamma)\epsilon=0 \, . \ee 
Here
$\Gamma$ is a generator of the local $\kappa$-symmetry. This gives
the following equation defining the BPS state 
\bea e^{ \left\{
-{1\over 2} \sigma_3 \otimes H^{1/2}(\vec y^2)
\Gamma_{67}[(\theta_1 + \theta_2) (1-\Gamma_{6789})+ (\theta_1 -
\theta_2) (1+\Gamma_{6789})]\right\} }\otimes \Gamma_{6789} \,
\epsilon_0 =\epsilon_0 \label{killing1} \ .\eea 
Here the fluxes on
D7 are $
 {\cal F}_{67}=\tan{\theta_1}$ and ${\cal F}_{89} = \tan{\theta_2}$.
 If
$\theta_1 = \theta_2=\theta$ we find 
\bea \exp\{-\sigma_3 \otimes
H^{1/2} \theta \, \Gamma_{67}(1- \Gamma_{6789})\}   \otimes
\Gamma_{6789} \, \epsilon_0 =\epsilon_0 \label{killing2}\ , \eea
where $\epsilon_0$ is a constant spinor. Thus for self-dual ${\cal
F}$ (${\cal F}^-=0$), an  exact non-linear Killing condition can
be satisfied subject to two projector equations on the Killing
spinors, $ \epsilon = i \sigma_2 \otimes \Gamma_{01236789}\epsilon
$ and $\epsilon =    \Gamma_{6789}\epsilon $. However, if
$\theta_1 \neq \theta_2$, equation (\ref{killing1}) cannot be
satisfied for non-vanishing $\epsilon_0$. Thus any small deviation
from the condition ${\cal F}^-=0$ leads to a  breaking of
supersymmetry.

We would like to find an effective theory language which will
describe the case of unbroken supersymmetry (with ${\cal F}^-=0$)
when the position of the D7 brane in a background of the D3 is a
modulus. Afterwards we will consider a deviation from the
supersymmetry by looking at the case ${\cal F}^-\neq 0$ and try to
keep the field describing the distance between branes as a
modulus.

We will try to construct an F-term potential
which coincides with the (AdS part of) KKLT model when the distance
between the branes is exactly zero. We will call a specific
field $S$ the ``inflaton'' in anticipation of it behaving
as such once the system has been uplifted to dS space. In the
AdS stage this field should just be a flat direction of the
potential. This means that we need a condition for unbroken
supersymmetry to be satisfied in the AdS space.

We assume that the \Ka\ potentials for the volume field $\rho$ and
the inflaton field $S$ are decoupled. This is different from the
corresponding assumption in the case of the D3 brane in
\cite{Kachru:2003sx} where the coordinate of the moving D3 brane
was related to the inflaton. Thus we assume that the \Ka\ potential is the one in eq. (\ref{kahlerD7}) and
that the superpotential, $W (\rho, S)$, is such that, at $S=0$, it
has a minimum in $\rho$, e.g. as in the KKLT model \be W(\rho,
S=0) = W_{KKLT}(\rho) \equiv w(\rho) \, .\label{KKLT}\ee For
example, the simplest $w(\rho)=W_{KKLT}(\rho) = W_0+Ae^{-a\rho}$.
Since we would like to describe a BPS state of branes, it is
natural to require that a condition of an exact unbroken
supersymmetry, 
 \be D_\rho W=0 \, ,  \qquad D_S W =0\, ,
\label{susy}\ee 
is satisfied at a particular value of
$\rho=\rho_{cr}$ and at all values of the separation
$s=\textrm{Re} S$. An exact solution of equations
(\ref{KKLT}) and (\ref{susy}) is available under the condition
that the supersymmetric extremum takes place at 
\be \rho=\bar \rho
\, , \qquad S=\bar S\, , \ee 
which for $ \rho=\sigma+i \alpha $ and 
$S=s+i\beta $ means that the state with unbroken supersymmetry
requires 
\be \alpha=0 \, , \qquad \beta=0 \, . \ee 
The solution
for the superpotential is given by the product of the KKLT
superpotential times an $S^2$-dependent exponent. 
\be W(\rho, S)=
W_{KKLT}(\rho)\, e^{-{S^2\over 2} }=w(\rho)\, e^{-{S^2\over 2} }\ .
\ee 
This leads to a simple potential which depends  only on
$\beta$ and does not depend on $s$. \be V^F(\rho, \bar \rho, S,
\bar S)={e^{-{1\over 2}(S-\bar S)^2} \over (\rho+\bar
\rho)^3}[|D_{\rho} w(\rho)|^2{(\rho + \bar{\rho})^2 \over 3} -
 (3 +(S-\bar S)^2) | w(\rho)|^2]\ .
\label{VF}\ee

Note that the field $S$ has a canonical kinetic term due the
presence of the $S\bar S$-term in the \Ka\ potential. Also, the
term $e^K |D_S W|^2$ is given by ${e^{-{1\over 2}(S-\bar S)^2}
\over (\rho+\bar \rho)^3}[
  -(S-\bar S)^2 | w(\rho)|^2]$.
At $S=0$ the potential coincides with the F-term potential of the KKLT model,
\be
V(\rho, S=0, \bar \rho, \bar S=0)= V^F_{KKLT}= {1\over (\rho+\bar \rho)^3}[|D_{\rho} w(\rho)|^2{(\rho + \bar{\rho})^2 \over 3}
 -3 | w(\rho)|^2]\ .
 \ee
In fact, the perfect flatness of the potential (\ref{VF}) in the
direction of the $s$ field was originally discovered by
plotting the potential as a function of $(\sigma, s)$ and
observing that the $s$-direction forms a trench.  When the
trench is cut at any constant value of
$s$ it gives the profile of the KKLT superpotential for the
volume $\sigma$.
\begin{figure}[h!]
\centering \epsfysize=8cm
\includegraphics[scale=0.9]{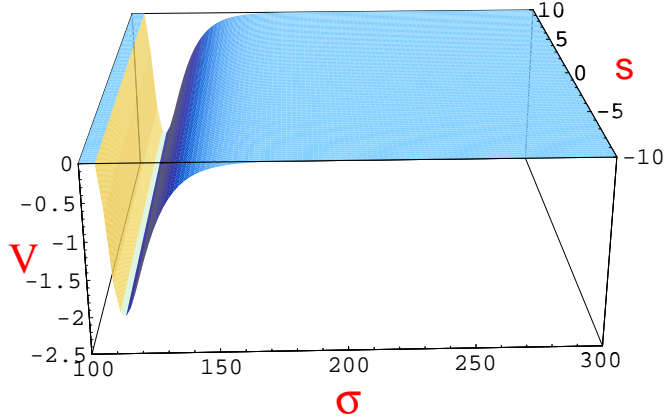}
\caption{AdS trench at the point in the moduli space where
$\alpha=0$ and $\beta=0$ which is a saddle point in these two
variables. The exact supersymmetric trench corresponds to the
minimum of the potential in the $\sigma$-direction and shows the
neutral equilibrium in the $s$-direction.} \label{AdS}
\end{figure}
The plot suggests that it may be possible to bring the potential
to a form that is manifestly independent of $s$, as one can see
in eq.(\ref{VF}). 

Another
way to see the manifest $s$ independence is to note that the
model can be described by a new $\tilde K$ and $\tilde W$ using
\Ka\ transformations so that 
\be \tilde K= -3\ln (\rho+\bar \rho)
-{(S-\bar S)^2\over 2} \, , \qquad \tilde W(\rho,S)= w(\rho) \,
.\label{newkahler}\ee 
The kinetic term for the $S$ field remains canonical. The
conditions for unbroken supersymmetry in new variables are 
\be
\tilde D_\rho \tilde W=D_\rho w(\rho)=0 \, , \qquad \tilde D_S
\tilde W= -(S-\bar S)\, w(\rho)=0 \, . \label{newkahler2}\ee 
These conditions are
satisfied at $\alpha=\beta=0$, i. e. at $\rho-\bar \rho=S-\bar
S=0$.

Thus we have presented here an effective model which may be
interpreted as describing some BPS branes in a compact space with
stabilized volume. The fact that the corresponding potential has a
flat direction $s$ follows from the requirement of the
existence in the theory of a supersymmetric critical point. The
potential has a flat direction describing the distance between
branes while keeping the volume fixed. 

In the language of the
$\tilde K$  and $\tilde W$, the existence of the supersymmetric flat direction is
a consequence of maintaining a shift symmetry of the model under
compactification: 
\be S \Rightarrow S+\delta \, , \qquad \tilde K
(S-\bar S) \Rightarrow \tilde K (S-\bar S) \, ,\qquad \tilde W (\rho)
\Rightarrow \tilde W (\rho)\ \, , \ee
where $\delta\in \mathbb{R}$\,. Clearly, the effective potential derived from $\tilde K$  and $\tilde W$ (\ref{newkahler}) does not have any dependence on $S+\bar S$, i.e. on the inflaton field $s=(S+\bar S)/2$. In fact, we could simply start with the effective $N=1$ supergravity (\ref{newkahler}) and use shift symmetry to prove the flatness of the potential in the inflaton direction. This would be reminiscent of the proposal in
\cite{Kawasaki:2000yn}, where the requirement of an analogous shift
symmetry was used to rescue chaotic inflation in supergravity.

An important difference in our setting is that we derived the condition of shift symmetry
of the effective $N=1$ supergravity as a consequence of our requirement to  describe the BPS branes 
using effective field theory \cite{Dasgupta:2002ew,Bergshoeff:1997kr}. That is why we needed to check the consistency condition (\ref{newkahler2}) in order to verify that the supersymmetric state corresponding to BPS branes does exist in our theory. However, once this property of the theory is established, we can use shift symmetry to prove that the potential remains flat in the inflaton direction, $s$, not only in the supersymmetric state in AdS, but for all values of the fields $\sigma$, $\alpha$ and $\beta$.

It is  instructive to present the potential
as an explicit function of all 3 remaining fields. Using the
example from KKLT where $w(\rho) = W_0+ Ae^{-a\rho}$ we find for $
V(\sigma, \alpha, \beta)$ the following expression:
\be {e^{2\beta^2-2 a\sigma}\over 6\sigma^3}
\left[A^2[a\sigma(3+a\sigma) +3\beta^2]+ 3A e^{a\sigma}(a\sigma
+2\beta^2)\cos(a\alpha) W_0+ 3 e^{2a\sigma}\beta^2 W_0^2 \right].
\ee 
The dependence on $\alpha$ is very simple: 
\be V(\sigma,
\alpha, \beta)= f(\sigma, \beta) + g(\sigma, \beta)
\cos(a\alpha)\, , \ee 
where $g(\sigma, \beta)$ is negative for all
values of $(\sigma, \beta)$ because $W_0$ is negative in this model.
This shows that, at all values of $\sigma$ and $\beta$, the
potential has a minimum at $a\alpha=2n \pi $ and a maximum at
$a\alpha=(2n+1) \pi $. Thus $\alpha=0$ is in fact a
supersymmetric minimum. From now on we will simplify our problem
by taking $\alpha=0$. 
\be V(\sigma, \alpha=0,
\beta)={e^{2\beta^2-2 a\sigma}\over 6\sigma^3}
\left[A^2[a\sigma(3+a\sigma) +3\beta^2]+ 3A e^{a\sigma}(a\sigma
+2\beta^2) W_0+ 3 e^{2a\sigma}\beta^2 W_0^2 \right]. \ee 
At the
critical point for $\sigma$ where $W_0=- Ae^{-a\,\sigma_{cr}}(1+
{2\over 3}\, a\, \sigma_{cr})$ the dependence on $\beta$
simplifies and we find 
\be V(\sigma_{cr}, \alpha=0, \beta)={a^2
A^2 e^{2\beta^2-2 a\sigma_{cr}}\over 18\sigma_{cr}} (-3+4\beta^2).
\ee 
The potential has, in addition to the supersymmetric critical
point at  $\beta=0$, another non-supersymmetric minimum at
$\beta^2=1/4$ and $D_\rho W(\rho, S)=0 \, ,  D_S W (\rho, S)\neq
0$. One finds that 
\be {\partial V(\sigma_{cr}, \beta)\over
\partial \beta}= 0 \ee when \be D_\rho w(\rho_{cr})=0 \, , \qquad
\beta (-1+ 4\beta^2)=0 \, .\ee 
The 
point $\beta=0$, $\alpha=0$, $\sigma=\sigma_{cr}$ is a
supersymmetric saddle point, which is a maximum in the $\beta$
direction and a minimum in the $\alpha$ and $\sigma$ directions.
The  
point $\beta=\pm 1/2$ and $\alpha=0$, $\sigma=\sigma_{cr}$ is a
non-supersymmetric minimum of the potential.

In our model the field $S$ is proportional to $x^4+ ix^5$, where the distance between branes is $d^2=(x^4)^2+ (x^5)^2$.  We see that the exact supersymmetric state is realized when the position of the  D7 coincides with the position of the stack of D3's in $x^5$ direction.  In such a case the distance in the direction of $x^4$, and therefore the total distance $d^2$, is completely arbitrary. It does not cost any energy to move the D7 in $x^4$ direction, which plays the role of the inflaton field in our model.

As in the KKLT model, the AdS part is just a technical step
towards the cosmological theory. The next step is to add a D-term
potential related to a D7 brane with the non-self-dual flux on its
worldvolume to lift the system to dS.

\section{dS uplifting}

The procedure of uplifting the AdS minimum to a dS one using an
anti-D3 brane was explained in \cite{Kachru:2003aw}. An
alternative method based on non-self-dual fluxes on D7 branes was
suggested in \cite{Burgess:2003ic}. For our purpose of
generalizing D-term inflation to the case of stabilized volume
compactification models, it is appropriate to use this alternative
method. In particular, this will allow us to study the exit from
inflation towards Minkowski or dS with a very small cosmological
constant. This will involve a few new steps: considering D3-D7
strings stretched between D3 and D7 branes ($\Phi_{\pm}$ fields)
and their logarithmic effect on the flat direction of the
classical potential. We will also be able to consider the
waterfall stage of the hybrid inflation model where $S$ is close
to zero and one of the charged fields, e.g. $\Phi_+$, will change
significantly.

When we are interested only in the inflationary stage of D-term
inflation with volume stabilization, the charged fields
$\Phi_{\pm}$ are near the dS minimum  with $\Phi_{+}=\Phi_{-}=0 $. At
$\Phi_{+}=\Phi_{-}=0 $  the D-term is
given by \cite{Burgess:2003ic} 
\be {g^2 \over 2} D^2 = { C\over
\sigma^3} \ .\ee 
Here $g^2$ for the vector fields on the D7 is
proportional to ${1\over \rho+\bar \rho}$  and the $\rho$-dependent $D$-term  is calculated via fluxes in the $6,7,8,9$
space of the D7 and is proportional to ${1\over \rho+\bar \rho}$.

The total classical potential describing the inflationary stage of
inflation with vanishing $\Phi_{\pm}$-fields consists of both
$V^F$ and $V^D$ and is given by 
\bea &&V(\rho, S, \bar \rho, \bar
S) = V^F(\sigma, \alpha, s, \beta)+ V^D(\sigma) ={e^{2\beta^2-2
a\sigma}\over 6\sigma^3}A^2[a\sigma(3+a\sigma) +3\beta^2]+
\nonumber\\
&&{e^{2\beta^2-2 a\sigma}\over 2\sigma^3} \left[ A
e^{a\sigma}(a\sigma +2\beta^2)\cos(a\alpha) W_0+
e^{2a\sigma}\beta^2 W_0^2 \right]+{ C\over \sigma^3} \ . \eea 
As
before, the $s$ direction is a flat one. We will consider a simple
case with $\alpha=0$ and $\beta=1/2$, i.e.  the
nonsupersymmetric minimum of $V$. The dependence of $V$
on the volume, $\sigma$, and the flat direction, $s$, is shown
in Figure 2.

\begin{figure}[h!]
\centering \epsfysize=8cm
\includegraphics[scale=0.9]{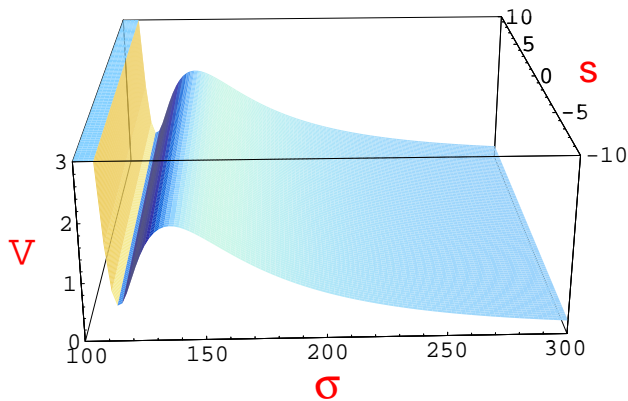}
\caption{dS trench at the minimum of $\alpha=0$ and $ \beta=1/2$
in which the volume $\sigma$ is stabilized close to the critical
value of the AdS stage. There is no dependence of $V$ on the inflaton
$s$. } \label{dS}
\end{figure}

Thus the volume is stabilized and the inflaton field (the distance
between the branes) can change without affecting the stabilization
mechanism. 
We are led at a picture which is not exactly D-term
inflation since we have a non-vanishing $V^F$-term in the potential.
However, the main features of the D-term inflationary model and
the corresponding D3/D7 brane construction  \cite{Dasgupta:2002ew} are preserved:
the D3-D7 strings stretched between branes (which correspond to the fields $\Phi_{\pm}$) have a mass
splitting due to a non-self-dual flux. This  leads to the shift symmetry breaking by quantum effects and to the logarithmic
corrections to the potential $\sim \log s$, which will drive
 hybrid inflation.

Here again, the effective field theory tells us that if the distance between   D7 from the stack of D3's in direction $x^5$ is defined by $x^5\sim \beta=1/2$,  it will stay there as it is a minimum of the potential. The position of D7 in the $x^4$ direction is almost flat: the classical potential is exactly flat\footnote{Note that until one considers the interaction of the inflaton field $s$ with the fields $\Phi_{\pm}$, the flatness of the effective potential is preserved even at the quantum level, despite supersymmetry breaking, because the masses of all other fields do not depend on $s$. We are grateful to A. Linde for the discussion of this issue.} but the first loop corrections $\sim \ln s$ trigger slow-roll inflation as in D-term inflation.

After passing the
critical distance, the D7 brane will form a supersymmetric bound
state with the D3 and one of the charged  fields $\Phi_+$ or
$\Phi_-$ will acquire a vev. Preliminary investigation shows that 
the volume of internal space will remain stabilized during the rapid waterfall stage at the end of inflation. We plan to perform 
%the numerical evaluations
a detailed investigation of this
model in future.

\section{Discussion}

In this note we proposed an ``inflaton trench'' mechanism which
may lead to a viable model of inflation with
volume stabilization in string theory. The shift symmetry of the
classical theory protects the small mass of the inflaton
in the framework of a string compactification with a stabilized
volume modulus. This symmetry is shown to follow from the
description of BPS branes in a supersymmetric equilibrium. A small
deviation from supersymmetry leads to a small violation of the
shift symmetry which allows a slow-roll regime of inflation. A similar mechanism may be responsible for the present stage of acceleration of the universe.

A picture of a ``stringy landscape'' that has been developed in
\cite{Susskind:2003kw} takes into account, in particular, a
profile of the KKLT model. Now it may be enriched  with the
trenches of the kind shown in Fig. 2  where inflation or late-time acceleration of the universe
may take place.

We are planning to develop these considerations in the context of the D3/D7
brane construction, with a hope that it may provide us with a realistic model of brane
inflation with stable compactification. 
One can also expect that the ``inflaton trench'' mechanism may be useful for other models of brane inflation. We hope to return to the discussion of these possibilities in a separate publication.

\subsection*{Acknowledgments}
It is a pleasure to thank P. Binetruy,  O. DeWolfe, M. Douglas, G. Dvali,  S. Giddings, S. Kachru, A. Linde,
J. Maldacena,  L. McAllister and S. Trivedi for useful discussions.  This work
was supported by NSF grant PHY-0244728.

\end{document}